\documentclass[
 reprint,
 superscriptaddress,
 nofootinbib,
 amsmath,
 amssymb,
 aps,
 pra,
 showkeys
]{revtex4-2}

\usepackage{graphicx}
\usepackage{xcolor}
\usepackage{dcolumn}
\usepackage{bm}
\usepackage{hyperref}
\hypersetup{
    colorlinks=true,
    linkcolor=blue,
    citecolor=blue,
    urlcolor=blue
    }

\newcommand{\Sr}{${}^{87}$Sr}
\newcommand{\Yb}{${}^{171}$Yb}
\renewcommand{\vec}{\boldsymbol}
\newcommand{\op}[1]{\hat{\vec{#1}}}
\newcommand{\ket}[1]{\left|#1\right\rangle}
\newcommand{\bra}[1]{\left\langle#1\right|}

\begin{document}

\title{Engineering Rydberg-pair interactions \\ in divalent atoms with hyperfine-split ionization thresholds}

\author{Frederic Hummel}
\email{fhummel@atom-computing.com}
\affiliation{Atom Computing, Inc., Berkeley, California}

\author{Sebastian Weber}
\email{weber@itp3.uni-stuttgart.de}
\affiliation{Institute for Theoretical Physics III and Center for Integrated Quantum Science and Technology, Universität Stuttgart, Germany}

\author{Johannes Mögerle}
\affiliation{Institute for Theoretical Physics III and Center for Integrated Quantum Science and Technology, Universität Stuttgart, Germany}

\author{Henri Menke}
\affiliation{Department of Physics, Friedrich-Alexander-Universität Erlangen-Nürnberg, Germany}

\author{Jonathan King}
\affiliation{Atom Computing, Inc., Berkeley, California}

\author{Benjamin Bloom}
\affiliation{Atom Computing, Inc., Berkeley, California}
 
\author{Sebastian Hofferberth}
\email{hofferberth@iap.uni-bonn.de}
\affiliation{Institute of Applied Physics, University of Bonn, Germany}

\author{Ming Li}
\email{mli@atom-computing.com}
\affiliation{Atom Computing, Inc., Berkeley, California}

\date{\today}

\begin{abstract}
Quantum information processing with neutral atoms relies on Rydberg excitation for entanglement generation.
While the use of heavy divalent or open-shell elements, such as strontium or ytterbium, has benefits due to their optically active core and a variety of possible qubit encodings, their Rydberg structure is generally complex.
For some isotopes in particular, hyperfine interactions are relevant even for highly excited electronic states.
We employ multi-channel quantum defect theory to infer the Rydberg structure of isotopes with non-zero nuclear spin and perform non-perturbative Rydberg-pair interaction calculations.
We find that due to the high level density and sensitivities to external fields, experimental parameters must be precisely controlled.
Specifically in \Sr, we study an intrinsic Förster resonance, unique to divalent atoms with hyperfine-split thresholds, which simultaneously provides line stability with respect to external field fluctuations and enhanced long-range interactions.
Additionally, we provide parameters for pair states that can be effectively described by single-channel Rydberg series.
The explored pair states provide exciting opportunities for applications in the blockade regime as well as for more exotic long-range interactions such as largely flat, distance-independent potentials.
\end{abstract}

\maketitle


\section{Introduction \label{sec:intro}}

The long-range interactions between Rydberg atoms cause exotic processes in cold quantum gases and enable applications in non-linear quantum optics \cite{Pritchard2010, Gorshkov2011, Firstenberg2016}.
Coherent Rydberg excitations are essential to quantum information processing using neutral atoms in order to introduce and mediate entanglement \cite{Saffman2010, Adams2020, Morgado2021}.
Ensembles of individually trapped atoms in optical tweezer arrays have enabled incredible progress for quantum simulation and computation with defect-free assembly \cite{Endres2016, Manetsch2024}, reconfigurable lattice geometries \cite{Barredo2016, Kim2016, Bernien2017, Ebadi2021}, mid-circuit erasure conversion and readout \cite{Ma2023, Norcia2023}, continuous reloading \cite{Tao2023, Norcia2024}, and unseen connectivities by aid of rearrangement \cite{Bluvstein2022}.
Recently, a variety of atomic species has been explored for quantum gas experiments and quantum technology applications exploiting different aspects of the atomic structure \cite{Browaeys2016, Jenkins2022, Chen2022, Unnikrishnan2024, Cao2024}.
While alkali metals benefit from a long history of scientific research, the presence of a second valence electron in alkaline-earth and some lanthanide elements provides compelling control opportunities \cite{Dunning2016}, e.g.~the core electron can be utilized to match polarizabilities of ground- and Rydberg states for magic wavelength trapping \cite{Wilson2022} and isolated core excitations can be exploited to introduce coherent light shifts and probe Rydberg states \cite{Burgers2022, Pham2022}.
Some isotopes allow for the storage of quantum information in their nuclear spin, achieving coherence times on the order of tens of seconds \cite{Barnes2022}.

Most schemes to entangle neutral atoms rely on strong dipolar interactions between Rydberg states, which are either coupled by near-resonant laser fields \cite{Jaksch2000, Lukin2001, Levine2019, Jandura2022} or by dressing ground states with far off-resonant lasers \cite{Bouchoule2002, Johnson2010, Pupillo2010, Mitra2020}.
Yet, for the Rydberg structure, typically only one or at most a few energy levels are considered explicitly.
Non-perturbative Rydberg-pair interaction calculations are required to confirm the absence of other pair states and leakage channels in the Rydberg spectrum \cite{Browaeys2016, Leseleuc2018, Weber2018}.

In this article, we present calculations of non-perturbative Rydberg-pair interactions for divalent atoms with hyperfine-split ionization thresholds in order to study interaction tunabilty and the robustness of perturbative predictions.
The complex spectra of these atoms are described within \emph{multi-channel quantum defect theory} \cite{Aymar1996}, which has been adapted to include nuclear spin using \emph{frame transformation} \cite{Robicheaux2018}.
Recently, \Sr~was predicted to feature unusual properties not encountered in atoms with a simpler structure, namely, for some Rydberg states, simultaneous insensitivity to electric fields and resonant dipolar interaction in equal-state pairs \cite{Robicheaux2019}.
We find this to be a general feature caused by the hyperfine interaction between different Rydberg series when the energy splitting of adjacent Rydberg manifolds becomes comparable to the threshold splitting in the ionic core.
Specifically, the coupling of Rydberg series belonging to different ionic-core hyperfine configurations gives rise to varying electron spin character of some Rydberg states as a function of the principal quantum number \cite{Liao1980, Beigang1981} and to a corresponding energy dependence of their quantum defects.
The resulting properties are appealing for quantum technology applications and Rydberg gas experiments: 
High degrees of level stability help preparing Rydberg states that minimize amplitude and phase errors.
Intrinsic Förster resonances lead to strong dipole-dipole interactions, allowing for the engineering of exotic many-body Hamiltonians and for the implementation of two- and multi-qubit gates \cite{Petrosyan2017}.
These Förster resonances do not require careful tuning of external fields, which is often the case for atoms with a single valence electron \cite{Walker2008, Ryabtsev2010, Tretyakov2014, Beterov2015, Paris2016}, but tuning of the principal quantum number is in general sufficient as we will demonstrate.
Moreover, the Förster resonances occur for two atoms being in the same state, allowing for global addressing of the atoms via a single laser beam.

This article is structured as follows:
In Sec.~\ref{sec:theory}, we review the theory and methods, particularly the Hamiltonian for a pair of Rydberg atoms in Sec.~\ref{subsec:rydberg}.
After this general discussion of Rydberg interactions, we provide the methods specific to divalent Rydberg atoms in Sec.~\ref{subsec:mqdt}, where we outline multi-channel quantum defect theory and frame transformation.
Sec.~\ref{subsec:sr87} details \Sr, with comments on exact diagonalization in Sec.~\ref{subsec:basis} and on the developed software in Sec.~\ref{subsec:software}.
Sec.~\ref{sec:results} contains the results, discussing the single-atom spectral structure in Sec.~\ref{subsec:resonance} and introducing field-free pair-interaction potentials for the Förster resonant Rydberg series in Sec.~\ref{subsec:fieldfree}.
Sec.~\ref{subsec:magnetic} covers the response of single atoms and Rydberg pairs to magnetic fields, similarly for electric fields in Sec.~\ref{subsec:electric}.
A comparison to a different, single-channel Rydberg series is provided in Sec.~\ref{subsec:11/2}.
Finally, Sec.~\ref{sec:conclusion} contains the conclusion and an outlook.

\section{Theory and Methods \label{sec:theory}}

\subsection{Rydberg-pair Hamiltonian \label{subsec:rydberg}}

\begin{figure}
\centering
\includegraphics[width=0.48\textwidth]{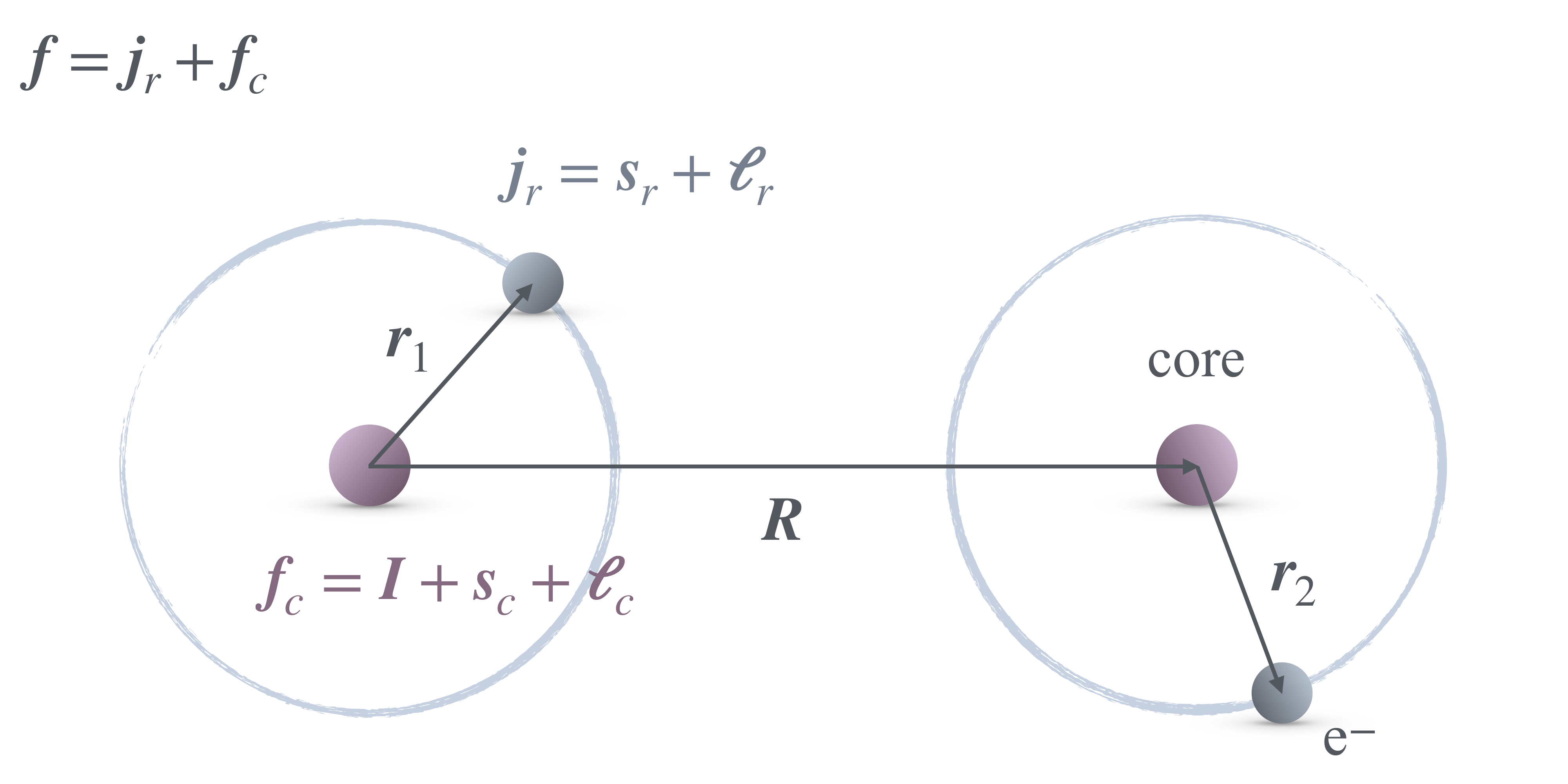}
\caption{\label{fig:scetch} 
The molecular system consisting of two Rydberg atoms with internuclear coordinate $\vec{R}$ and electronic coordinates $\vec{r}_i$ for atom $i$, respectively.
For each individual atom, the total hyperfine spin $f$ is a good quantum number.
In the fragmentation frame, $\vec{f}$ is obtained by coupling the ionic core's hyperfine angular momentum $\vec{f}_c$ and the Rydberg electron's total angular momentum $\vec{j}_r$.
The coupling scheme for short electron separations $\vec{r}_i$ depends on the particular eigenstate.
}
\end{figure}

We consider a pair of atoms as illustrated in Fig.~\ref{fig:scetch} and modelled by the Hamiltonian 
\begin{equation}
    H_2 = H_1 \otimes 1\!\!1 + 1\!\!1 \otimes H_1 + H_\text{int}(\vec{R}), \label{eq:pairHam}
\end{equation}
where $H_1$ is the single-atom Hamiltonian detailed further below and (using atomic units throughout)
\begin{equation}
    H_\text{int}(\vec{R}) = \sum_{\kappa_1, \kappa_2=1}^\infty V_{\kappa_1 \kappa_2} |\vec{R}|^{-1-\kappa_1-\kappa_2} \label{eq:multipole}
\end{equation}
is the multipole expansion of the electrostatic interaction between the two atoms suitable for internuclear separations exceeding the Le Roy radius and neglecting retardation effects.
Here, $V_{\kappa_1 \kappa_2}$ is proportional to the spherical multipole operators on the Rydberg electron in atom 1 and 2, respectively.
The exact form of $V_{\kappa_1 \kappa_2}$ depends on the choice of the coordinate system.
We choose the internuclear axis $\vec{R}$ to align with the $z$ axis and restrict Eq.~\ref{eq:multipole} to dipole-dipole interactions:
\begin{equation}
    V_{11} = -\sum_{q=-1}^1 \binom{2}{1+q} r_{1q}^{(1)}r_{1-q}^{(2)},
\end{equation}
where $r_{\kappa q}^{(i)}$ is the spherical multipole operator of order $\kappa$ for atom $i$.

We find a matrix representation of $H_2$ by expanding in a set of appropriately symmetrized wave functions of pairs of atoms.
In the absence of external fields, $H_2$ preserves reflection symmetry through a plane containing the internuclear axis.
Since we consider homonuclear pairs, $H_2$ also preserves inversion symmetry.
In the case of external field alignment with the internuclear axis, the total magnetic quantum number $M=m_1+m_2$ is a conserved quantity, where we introduce an index, indicating $m_i$ corresponds to atom $i$.
Finally, since only dipole-dipole interaction are considered in the multipole expansion (Eq.~\ref{eq:multipole}), $H_2$ preserves permutation symmetry.
We use these symmetries to decompose $H_2$ into blocks that can be treated independently of each other \cite{Weber2017}.

The Hamiltonian for a single Rydberg atom is given by
\begin{equation}
    H_1 = H_0 + H_\text{s} + H_\text{z} + H_\text{d},
\end{equation}
where $H_0$ is the field-free Hamiltonian with the good quantum numbers $f$ and $m$ corresponding to the total hyperfine spin, as well as parity.
The eigenvalues of $H_0$, 
\begin{equation}
    E_{\nu f} = E_I-\frac{R_\mu}{\nu^2}, 
\end{equation}
are conveniently expressed in terms of an effective principal quantum number $\nu$, defined with respect to the ionization threshold $E_I$.
Here, $R_\mu$ is the mass scaled Rydberg constant.
If an eigenstate is a superposition of multi-channel configurations as described in the following section, it is still useful to choose a specific $E_I$ in order to label eigenstates by a global $\nu$ and, hence, indicate their ordering.
For hyperfine split-ionization thresholds, we choose $E_I$ to be the energetically higher one, which some single-channel Rydberg series converge to.

The effect of static and homogeneous external fields is considered by the following three terms:
For the electric field $\vec F$, the Stark Hamiltonian is given by
\begin{equation}
    H_\text{s} = -\vec r \cdot \vec F,
\end{equation}
where $\vec r$ is the electronic coordinate (and, since the electron charge $e=1$ in atomic units, corresponds to the electric dipole moment).
For the magnetic field $\vec B$, the Zeeman Hamiltonian is given by
\begin{equation}
    H_\text{z} = -\vec \mu \cdot \vec B,
\end{equation}
where $\vec \mu$ is the magnetic dipole moment taking into account the electronic and nuclear spins
\begin{equation}
    \vec \mu = -\mu_B \left[ \vec{\ell}_c + \vec{\ell}_r + g_s \left( \vec{s}_c + \vec{s}_r \right) \right] + \mu_I \vec I,
\end{equation}   
where $\mu_B$ and $\mu_I$ are the Bohr magneton and the nuclear magnetic moment and $g_s$ is the electron spin g-factor.
Here, the subscripts, $c$ and $r$ label core and Rydberg electron quantum numbers, while $\vec \ell$ and $\vec s$ are orbital and electronic spin angular momentum, respectively, and $\vec I$ is the nuclear spin (cf.~Fig.~\ref{fig:scetch}).
The diamagnetic Hamiltonian is given by
\begin{equation}
    H_\text{d} = \frac{1}{8} \left( \vec{r}^2 \vec{B}^2 - \left| \vec r \cdot \vec B \right|^2 \right).
\end{equation}
This term is typically considered for ground-state atoms only when the magnetic field is on the order of 1 atomic unit ($\sim 10^4$ T).
However, as it scales with the surface area of the atom, $H_\text{d} \sim n^4$, for the Rydberg states considered here, diamagnetic interaction is comparable to Zeeman interaction for fields as low as $\sim 100$ G.

In practise, before dealing with the interactions between two atom, the total single-atom Hamiltonian $H_1$ is constructed within a restricted basis (see Sec.~\ref{subsec:basis}) and diagonalized for a specific external field configuration; a process we call \emph{prediagonalization} \cite{Weber2017}.
An appropriate two-atom basis is then constructed by pairwise, symmetrized combinations of the resulting eigenstates with the option to provide an energy cutoff.
This is particularly relevant in order to only include well converged single-atom states in the two-atom basis.
Finally, Rydberg pair-interaction potentials are obtained by constructing $H_2$ in the two-atom basis and subsequent diagonalization at the internuclear distances of interest.

\subsection{Multi-channel quantum defect theory \label{subsec:mqdt}}

In order to find the eigensystem of $H_0$, we follow the formalism presented by Robicheaux et al.~combining multi-channel quantum defect theory (MQDT) and frame transformations to capture complex spin interactions in heavy atoms \cite{Robicheaux2018}.
Generally, the perspective of quantum defect theory is to consider an atom as a collision of an electron and an ionic core \cite{Seaton1966}.
In MQDT, the collision is described in terms of the real, symmetric $K$ matrix that parameterizes the scattering between different spin configurations, so-called channels \cite{Aymar1996}.
The frame transformation is the process of unitarily transforming $K$ matrices according to different coupling schemes that depend on the relevant length scale of the Rydberg electronic coordinate \cite{Fano1970}.
Each frame has a different advantage: 
In the fragmentation frame, it is straight-forward to incorporate multiple ionization thresholds and account for atomic hyperfine structure (or in the case of molecules, for vibrational or rotational degrees of freedom).
The close-coupling frame, on the other hand, provides a diagonal representation of the K matrix.

From a scattering perspective, the total hyperfine spin of the Rydberg atom $f$ is formed by coupling the asymptotically separated spins of the Rydberg electron $j_r$ and the ionic core's hyperfine spin $f_c$.
However, at small separations, $j_r$ and $f_c$ are not meaningful and the fragmentation-frame coupling scheme has to be replaced by a more suitable close-coupling scheme, e.g.~spin-orbit coupling.
This provides the additional advantage that the the K matrix in the close-coupling frame, obtained from spectroscopic data, are applicable (to a good approximation) independently of the isotope and the corresponding nuclear spin configuration.
An intuitive explanation for this is provided by comparing the energy scales of different spin couplings: 
Close to the ground-state, the hyperfine structure is only a small correction to the eigenstates dominated by spin-orbit coupling (or $j_cj_r$-coupling for some excited states, $j_c$ being the total angular momentum of the core electrons) and therefore has very limited impact on the coupling parameters.

The formalism presented in \cite{Robicheaux2018} and outlined below solves for the bound states of $H_0$ after transforming from the closed-coupling frame into the fragmentation frame.
This is equivalent to, but leads to different equations than solving directly in the close-coupling frame, presented e.g.~in \cite{Robaux1982}.
Let us start in the close-coupling frame, in which we assume the $K$ matrix to be diagonal, denoted as $\bar{\vec{K}}$, with elements
\begin{equation}
    \bar{K}_\alpha = \tan \pi \mu_\alpha,
\end{equation}
where $\mu_\alpha$ are (possibly energy dependent) quantum defects and $\alpha$ labels the close-coupling channels.
We choose the transformation into the fragmentation frame to be expressed as a series of unitary matrices
\begin{equation}
    \op{U}_{cc-fj} = \op{U}_{jj-fj} \op{U}_{LS-jj} \op{U}_{cc-LS}, \label{eq:unitary}
\end{equation}
where the subscripts label coupling schemes (detailed below), the first two matrices are solely determined by the corresponding angular momentum recouplings.
$\op{U}_{cc-LS}$ accounts for corrections to pure spin-orbit ($LS$) coupling that is typically expressed as a series of simple rotations between specific $LS$ channels, the (possibly energy dependent) angles of which serve as an optimization parameter in order to fit experimental data\footnote{
For light atoms, this is often not required and the $K$ matrix is diagonal in the $LS$ frame.
As pointed out in \cite{Robicheaux2019}, for \Sr, a coupling between the $5snd$ series singlet and triplet states, ${}^1D_2$ and ${}^3D_2$, can be accounted for by this additional rotation.}.
We list here the representations of the coupling schemes that make for the frame transformation matrices analogously to references \cite{Robicheaux2018, Robicheaux2019}.
When the Rydberg electron is close to the core, we write the wave function as
\begin{equation}
    |LS\rangle = |(((s_cs_r)S(l_cl_r)L)JI)fm \rangle,
\end{equation}
where parentheses indicate which spins are coupled consecutively.
This configuration is recoupled in an intermediate step to
\begin{equation}
    |jj\rangle = |(((s_cl_c)j_c(s_rl_r)j_r)JI)fm \rangle,
\end{equation}
before the final recoupling to the fragmentation frame
\begin{equation}
    |fj\rangle = |(((s_cl_c)j_cI)f_c(s_rl_r)j_r)fm \rangle.
\end{equation}
This scheme can be summarized to be propagating the coupling of the valence electrons to the nuclear spin through different stages.

Finally, the $K$ matrix in the fragmentation frame is obtained as
\begin{equation}
    \op K = \op{U}_{cc-fj} \bar{\vec{K}} \op{U}_{cc-fj}^\dagger. \label{eq:uku}
\end{equation}
Generally, the $K$ matrix is block diagonal in $f$ and in parity.
Thus, under the assumption $\ell_c=0$, it further decomposes into blocks of equal orbital angular momentum $\ell_r$.

A bound state $|\psi_{\nu f}\rangle$ at energy $E_{\nu f}$ is obtained by solving
\begin{subequations}
    \label{eq:roots}
    \begin{gather} 
        \det \op M = 0, \\
        \op M = \tan \vec{\beta} \ \op{1\!\!1} + \op K, 
    \end{gather}
\end{subequations}
where the elements of $\vec \beta$, $\beta_i = \pi(\nu_i - \ell_{r,i})$, are related to the relative principal quantum numbers $\nu_i$ defined with respect to the associated ionization threshold of the $i$-th channel, i.e.
\begin{equation}
    E=E_i-\frac{R_\mu}{\nu_i^2}, \ \forall i, 
\end{equation}
for total energy $E$ and thresholds $E_i$.
At a bound state energy $E_{\nu f}$, the coefficients for each channel $A_i$ are defined via the one-dimensional null space
\begin{subequations}
    \label{eq:nullspace}
    \begin{gather} 
        \vec A = \ker \left( (\vec \nu^{-3/2} \cos\vec\beta)^\top . \op M \right), \\ 
        \vec A^2 = 1,
    \end{gather}
\end{subequations}
where the exponentiation of $\vec\nu$, containing elements $\nu_i$, as well as the matrix multiplication (.) are assumed to be element-wise operations.
Then, the total MQDT wave function is given as
\begin{equation}
    |\psi_{\nu f}\rangle = \sum_i A_i^{(\nu f)} P_{\nu_i\ell_{r,i}}(r) |\phi_i\rangle, \label{eq:totalwave}
\end{equation}
with the channel function $|\phi_i\rangle$ representing the ionic core as well as the spin and angular configuration of the Rydberg electron.
The radial representation of the Rydberg electron is given by
\begin{equation}
    P_{\nu_i\ell_{r,i}}(r) = \left( f_i \cos \beta_i + g_i \sin \beta_i \right) \nu_i^{-3/2}, \label{eq:radialwave}
\end{equation}
with the regular and irregular Coulomb functions $f_i$ and $g_i$, respectively.
Eq.~\ref{eq:roots} guarantees the Rydberg wave function to go to zero asymptotically.

The total wave functions (Eq.~\ref{eq:totalwave}) containing the radial representation only for the Rydberg electron (Eq.~\ref{eq:radialwave}), but angular representation for both the Rydberg electron and the ionic core, are used for the calculation of all matrix elements, relevant for constructing the Hamiltonians discussed in the previous sections, particularly the electric multipole moments of order $\kappa$
\begin{equation}
    \bra{\psi_{\nu f}}r_{\kappa q}\ket{\psi_{\nu'f'}} = \sum_{i,j} A_i^{(\nu f)} A_j^{(\nu' f')} R_{ij}^\kappa \bra{\phi_i}Y_{\kappa q}\ket{\phi_j},
\end{equation}
with the spherical harmonics $Y_{\kappa q}$ and the radial integrals
\begin{equation}
    R_{ij}^\kappa = \int \mathrm{d}r \ r^\kappa P_{\nu_i\ell_{r,i}}(r) P_{\nu'_j\ell'_{r,j}}(r),
\end{equation}
as well as the magnetic dipole moment
\begin{equation}
    \bra{\psi_{\nu f}}\vec \mu\ket{\psi_{\nu'f'}} = \sum_{i,j} A_i^{(\nu f)} A_j^{(\nu' f')} R_{ij}^{0} \bra{\phi_i} \vec \mu \ket{\phi_j}.
\end{equation}
This assumes the approximation that contributions from the dipole operators acting on the core are negligible compared to contributions from the Rydberg electron because the corresponding wave function differ vastly in size.
Here, the representation in the fragmentation channel is beneficial: Rydberg and core quantum numbers are well separated therein.

\subsection{Strontium 87 \label{subsec:sr87}}

For the Rydberg structure of \Sr, we use the experimentally reported quantum defects for ${}^{88}$Sr in the $LS$ frame as listed in \cite{Robicheaux2019}.
We are interested in the $5sn\ell$ Rydberg series close to the ionization threshold at relatively large principal quantum numbers $n\gtrapprox30$.
In Sr, some doubly excited states of the $5pnp$-, $4dnd$-, and $4dnp$-series lie energetically below the first ionization threshold and are particularly relevant in order to accurately describe the spectrum at low $n\lessapprox15$ \cite{Vaillant2014}.
However, for our purposes, the influence of these series is sufficiently captured by considering the quantum defects of the associated $5sn\ell$ series as slowly varying functions of energy \cite{Robicheaux2019}.
In the case of \Sr$^+$ with its fermionic core and neglecting electron excitations, i.e.~$\ell_c=0$, the finite nuclear spin $I=\frac{9}{2}$ couples to the electronic angular momentum $j_c=s_c=\frac{1}{2}$ giving rise to two core hyperfine configurations $f_c$ with energies separated by roughly 5 GHz.
The energies of the ionic ground states with different hyperfine configurations serve as distinct ionization thresholds $E_{f_c}$ for the Rydberg electron.

To visualize which pair-interaction potentials can possibly be excited via an intermediate level |c>, we use a color code highlighting the potentials according to the dipole coupling of Rydberg-pair states $\ket{\Psi}$ to the pair state $\ket{c,t}$, where $\ket{t}$ is the Rydberg target state to be specified and we have exemplary chosen $\ket{c}$ to be the ${}^3P_0$ $f=\frac{9}{2}$ clock state.
Thus, the coloring corresponds to the matrix element $D=|\bra{\Psi} \vec{r} \otimes 1\!\!1\ket{c, t}|$.
Note that the state $\ket{c, t}$ is not symmetrized, therefore the coloring is not sensitive to being in the dark- or bright-state sector of the Hamiltonian.
In pure Born-Oppenheimer approximation, these sectors are well separated, however, considering motional degrees of freedom, there is some finite coupling.
We consider this choice for coloration to be more instructive than coloring according to the overlap with a specified target state, because the latter neglects the impact of other dipole coupled states with for example differing hyperfine configuration or different orbital angular momentum.
For e.g.~blockade experiments, it is crucial to find an energy window as well as a region of internuclear distances, which is ``clean'', i.e.~free of dipole-coupled potentials.
Unless otherwise specified, we choose for all pair-interaction potential figures in this article, $D>10^{-5}\ ea_0$ for colored points and $D>10^{-9}\ ea_0$ for gray, high-opacity points to inform about the encountered level densities.

\subsection{Remarks on basis sets \label{subsec:basis}}

A major challenge for the exact diagonalization of Rydberg pair Hamiltonians is the choice of the basis set and the corresponding convergence of eigenstates.
For van-der-Waals type interaction, the pair states relevant to converge a given pair potential can be energetically distant, defining a requirement for the considered energy window for diagonalization.
In general external field configurations, when $M$ is not a conserved quantity, the pair state level density becomes large and the number of required basis states is typically the limiting factor.
Adding to this precarity, for multi-channel atoms, the single-atom level density is generally larger than e.g.~in alkali atoms due to the relevance of the hyperfine interaction.

To illustrate this, let us consider the number of bound states within a single Rydberg manifold restricting the maximal orbital angular momentum to $\ell_\text{max}=3$.
In ${}^{87}$Rb, where $I=\frac{3}{2}$, there are technically 32 levels with average degeneracy of 4 each.
However, since the hyperfine interaction scales as $n^{-3}$, the nuclear spin can be disregarded for sufficiently large principal quantum numbers shrinking the number of levels to 7, or 32 in the case of resolved degeneracy.
Contrary for \Sr, hyperfine interaction cannot be disregarded\footnote{
In alkali metals, the hyperfine interaction scales with the principal quantum number due to the overlap of the electron wave function with the nuclear core.
While in alkaline-earths the Rydberg electron is similarly disconnected from the nucleus, the second valence electron is always close to the core probing its structure.
The electron-spin coupling of the two valence electrons prevents decoupling of the hyperfine structure and the corresponding scaling.
}
and the number of levels is 64 with average degeneracy 10 each, due to the large nuclear spin $I=\frac{9}{2}$.
A factor of 20 in the single atom level density has devastating consequences for the diagonalization of pair states, where the number of states scales quadratically and the number of matrix elements to construct the Hamiltonian scales with the power of 4.

In order to achieve convergence on the sub-MHz scale for the figures in this article, the typical basis sizes range from 10k to 25k states, after symmetrization.
For external fields not aligned with the $z$ axis (not discussed here), we were only able to converge potentials to the tens-of-MHz scale using around 45k basis states.
Unless otherwise specified, we take into account 3 to 4 neighboring $n$ manifolds, i.e.~7 to 9 manifolds in total, and orbital angular momenta $\ell$ from 0 to 3 in the single atom basis.
States further separated from the target state typically do not have the coupling strength to significantly contribute, although even energetically distant states could, in principle, make a degenerate pair.

For calculations with finite electric field, we compare the results including higher $\ell$ up to 4 or 5, however this does not serve as a proper convergence check in the case of high electric fields, because contributions from degenerate high-$\ell$ states could, in principle, add up to modify the level structure, hence, the focus on small electric fields, where we expect those contributions to be small.
Combinations of such single atom eigenstates are considered in the pair basis, if their total energy is within a specified energy window of the target state.
We found 20 GHz to be sufficient for our convergence goal for the considered states with $n\approx60$, however, in arbitrary external field configurations, this may lead to intractable basis sizes and the convergence criterion may need to be loosened.

\subsection{Pairinteraction software \label{subsec:software}}

In order to account for the increased requirements associated with calculating pair-interaction potentials of multi-channel atoms discussed in the previous sections, we extended the \emph{pairinteraction} software.
A corresponding open-source release is currently pending.\footnote{
We refer to the software release and the accompanying publication for a more complete description of the improvements and the corresponding capabilities.
The software will be available at \url{https://pairinteraction.github.io/}.}
Notably, the software supports sophisticated diagonalization routines such as the \emph{FEAST} algorithm \cite{Polizzi2009}.
It is planned to offer cloud access to the \emph{pairinteraction} software making use of CUDA for distributed computation on GPUs \cite{cuda}.
Furthermore, we propose a database standard for atomic state data.
Such a database contains the bound state energies and appropriate state labeling according to the relevant quantum numbers as well as, importantly, the reduced matrix elements required for constructing Hamiltonians with high-performance \cite{duckdb}.

Such a database separates the calculation of pair-interaction potentials from the more fundamental calculation of atomic spectra and, particularly, removes any requirement for the software to handle MQDT wave functions.
Instead, we provide a \emph{Julia} \cite{Julia2017} implementation of MQDT using frame transformations following Robicheaux et al.~\cite{Robicheaux2018} and outlined above as a utility tool for \emph{pairinteraction} that will be available as open-source software.
An MQDT spectrum is generally sensitive on the underlying quantum defect data and the considered channel couplings.
Therefore any experimental advances in the spectroscopy of multi-channel atoms and improved efforts to fit quantum defects will, in principle, lead to new databases that can be used with the \emph{pairinteraction} software.

\section{Results \label{sec:results}}

In the following we will show that the multi-channel couplings stemming from the hyperfine splitting of the ionic core and resulting in singlet-triplet mixed Rydberg series extends the complexity and versatility of Rydberg pair interactions.
Specifically, we focus on an intrinsic equal-state Förster resonance in the $5sns$ $f=\frac{9}{2}$ series of \Sr.
We can control the corresponding Rydberg-pair interactions in four ways, namely by the choice of the principal quantum number, by the magnitude of homogeneous magnetic and electric fields, and by the internuclear distance between atoms.
In contrast, the $5sns$ $f=\frac{11}{2}$ series of \Sr~is effectively described in a single channel approximation and provides the features known from Rydberg states in alkali metals with properties regularly scaling with the principal quantum number.

\subsection{Overview of the Förster resonance \label{subsec:resonance}}

\begin{figure}
\centering
\includegraphics[width=0.48\textwidth]{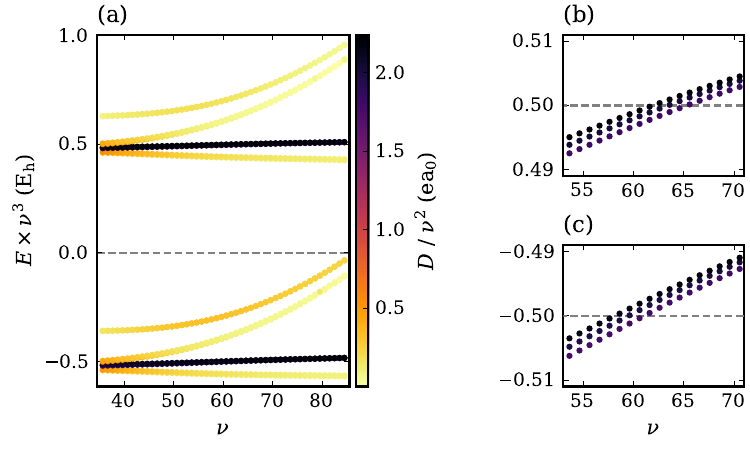}
\caption{\label{fig:levels} 
Transition energies to the $5snp$ manifold and reduced dipole matrix elements for the mixed singlet-triplet series $5sns$ $f=\frac{9}{2}$.
Each Rydberg manifold provides a total of ten different available states (colored points, some on top of each other), three of which are particularly strongly coupled.
Rydberg scaling of energies and dipole elements is taken into account and the $s$ state's energy is set to zero for reference.
(a) Each $p$ series has a characteristic energy dependence.
(b-c) Zooms into (a), the three strongly coupled states are equidistant from the $s$ state around $\nu=60$.
Each combination of these series gives rise to a resonant pair with slightly different Förster defect leading to complex Rydberg-pair interaction behavior.
}
\end{figure}

Fig.~\ref{fig:levels} illustrates the level structure and coupling elements of \Sr~relevant for the composition of Rydberg pair-interaction potentials as functions of the effective principal quantum number $\nu$.
(For this series $n\approx\nu+3.5$, where $n$ counts the electronic orbital.)  
Accounting for the relevant scaling associated with these properties, we show the level separation of the $5sns$ $f=\frac{9}{2}$ series to its dipole-coupled states, i.e.~states of $5sn'p$ $f'=\{\frac{7}{2}, \frac{9}{2}, \frac{11}{2}\}$ character.
Those states determine both the polarizabilty and the character of the pair-interaction potentials.
Each Rydberg manifold provides in total ten relevant states (not all of which are resolved in the figure), that can be grouped according to their total electronic character $S$, which can be mixed or pure.
A subset of three states has very similar mixed singlet-triplet character to the $5sns$ $f=\frac{9}{2}$ series and their dipole coupling is correspondingly strong.

Importantly, the position of the states of neighboring manifolds within the spectrum relative to the $s$ series slowly changes as a function of $\nu$ and around $\nu\approx61$ the states of the neighboring manifolds are equidistant, cf.~Fig.~\ref{fig:levels}(b-c).
In other words, the Förster defects, i.e.~the energy differences, of pairs $(n-1)p+np$ and $ns+ns$ have zero crossings giving rise to a broad Förster resonance.
Additionally, the contribution of these transitions to the Stark effect cancel giving rise to near zero polarizabilty.
Both of these effects have been predicted by perturbative considerations \cite{Robicheaux2019}.
Fig.~\ref{fig:levels} reveals the complex interplay of the contributing states.
The more weakly coupled states as seen in Fig.~\ref{fig:levels}(a), which have strong triplet character, contribute to the overall structure and illustrate the significant energy dependence of quantum defects, which originates in the hyperfine-split ionization threshold of \Sr${}^+$.

In the following, we analyze the conditions for this Förster resonance, indicating that it is intrinsic to an equal-state pair in the $s$ series strongly coupling to the adjacent $p$ levels at the zero crossing of the associated Förster defect, which is unique to specific Rydberg series in divalent atoms.
In Rydberg series with slowly varying energy dependence of the quantum defects, for example the $5sns$ $f=\frac{11}{2}$ series in \Sr, the functional behavior of the Förster defect features the regular $n^{-3}$ scaling.
This scaling and the consequential absence of a zero crossing of the Förster defect prohibits the existence of a Förster resonance.
Particularly, this scaling is generally present in alkali atoms, where Förster resonances are not typically found in equal-state pairs and rarely without additional tuning of energy levels with external fields \cite{Walker2008}.
Instead, the existence of a Förster resonance requires the analysis of many spectral properties and the careful tuning of the quantum numbers to different values for the involved atoms in order to achieve degeneracy of a pair \cite{Beterov2015}.
Examples are $p$ series in which the fine structure sufficiently shifts the state to be resonant with adjacent $s$ states or an $nsn'd$ pair or inter-species resonances using different principal quantum numbers for each species \cite{Ireland2024}.

In hyperfine-split ionic-core species, however, the existence of a mixed single-triplet $s$ series featuring a Förster resonance is guaranteed, because the hyperfine interaction leads to an exchange in electronic character as a function of $n$, when the intrinsic splitting of adjacent Rydberg states is comparable to the ionic hyperfine splitting.
For example, we expect such a resonance for the $6sns$ $f=\frac{1}{2}$ series of \Yb, however, the uncertainty of quantum defects of the $5snp$ series makes a quantitative prediction of its location in the spectrum currently dispensable\footnote{
On the other hand, the quantum defects for the $s$ and $d$ series in Yb are rather well known.
It is, however, inherently difficult to probe the $p$ series in heavy divalent atoms, in which Rydberg spectroscopy is typically performed using two-photon transitions, because the single photon transition is too deeply ultra violet.
We hope to see new Yb spectroscopy data soon or that this article can serve as a motivation for additional experimental efforts.
}.

The unique combination of the features makes the Förster resonant state an interesting candidate for quantum technology applications:
The bare Rydberg line is largely insusceptible to external fields and their fluctuations.
Beyond the small polarizability, the mixed electron-spin character of the state also leads to a suppressed linear Zeeman splitting, because singlet states have a vanishing $g$ factor, and the response to magnetic field is mostly quadratic driven by diamagnetic interaction requiring large fields.
At the same time, the pair state is subject to dipole-dipole interactions that provide a larger blockade radius than typically encountered for pair states subject to induced-dipole (i.e.~van-der-Waals) interactions.
And finally, the equal-state pair can be easily addressed in a global beam setup requiring only a single wavelength as opposed to Förster states involving pairs of states differing in $n$ or $\ell$ or even interspecies resonances as recently demonstrated \cite{Anand2024}.

\subsection{Field-free pair-interaction potentials \label{subsec:fieldfree}}

\begin{figure*}
\raggedleft
\includegraphics[width=1.06\textwidth]{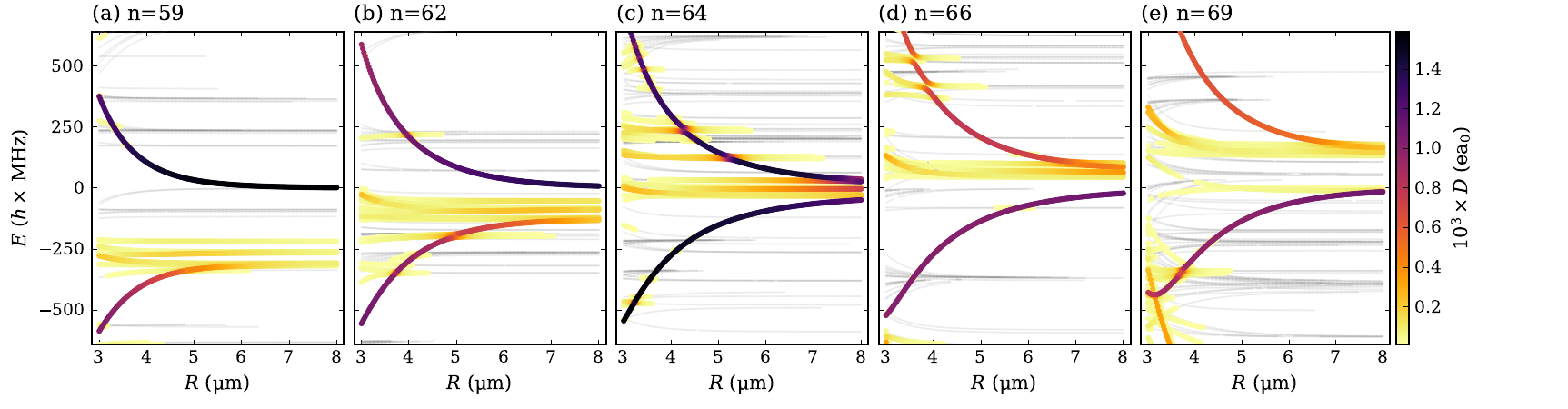}
\caption{\label{fig:forster} 
Pair-interaction potentials for the $5sns$ $f=\frac{9}{2}$ equal-state pairs of \Sr~for different $n=\{59, 62, 64, 66, 69\}$ (a-e).
Varying $n$, the potentials change character from repulsive to attractive, while on resonance (c), the potentials are symmetric and asymptotically degenerate.
However, many other pair states contribute to the potential energy landscape introducing avoided crossings and possible channels for population transfer.
The color indicates the dipole transition strength from the ${}^3P_0$ clock state.
Only the $M=-9$ symmetry sector is shown.
}
\end{figure*}

In Fig.~\ref{fig:forster} we show the Rydberg-pair interaction potentials for equal pairs in the $5sns$ $f=\frac{9}{2}$ series in the vicinity of the Förster resonance at $n=64$.
Here, the asymptotic energy of the $5sns + 5sns$ pair is set to zero and the coloring indicates dipolar coupling to the ${}^3P_0$ $f=9/2$ clock state.
Over a relatively short range of principal quantum numbers, $n=\{59, 62, 64, 66, 69\}$ shown in subfigures (a) through (e), the potentials drastically change character.
At low $n$ far from resonance, the potential is repulsive and scales as $R^{-6}$ with the internuclear distance, indicating induced-dipolar interactions, and the pair interaction is well described using perturbation theory.
Closer to resonance, the scaling changes to $R^{-3}$, indicating permanent-dipole interactions due to the mixing of pair states with $s$ and $p$ character.
At $n=64$, the $s$ and $p$ pairs are asymptotically degenerate and the molecular eigenstates are symmetric superpositions, which leads to symmetrically attractive and repulsive potentials as is characteristic for a Förster resonance.
However, the fact that several pairs of $p$ character contribute, leads to some non-generic level crossings and state mixings at large internuclear distances and at small detunings.
Furthermore, the significant $p$ state character of the potentials gives rise to yet more additional couplings to other pair states of either $s$ or $d$ character at larger energies (e.g.~around $E/h=150$ MHz and $E/h=250$ MHz) that lead to avoided crossings.
At larger $n$, the dominant potential remains attractive and the scaling transitions to dominantly $R^{-6}$ again.
At short internuclear distances $R<4$ µm, the potential is modified due to significant dipolar interactions stemming from other pair states at smaller energies.
The larger $n$, the more likely such modifications become, due to the increasing level density and the increasing dipolar interaction strength.

\subsection{Magnetic field response \label{subsec:magnetic}}

\begin{figure}
\raggedleft
\includegraphics[width=0.51\textwidth]{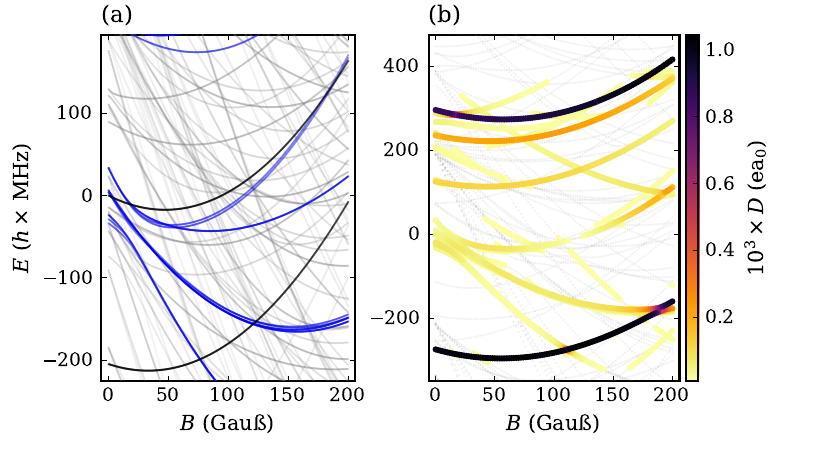}
\caption{\label{fig:magnetic}
Magnetic field maps for Rydberg pairs in the vicinity of the Förster resonant pair.
(a) Asymptotically, in the non-interacting limit, pair states are subject to vastly differing field responses all subject to diamagnetic interaction.
Black indicates pairs of $s$ states, blue pairs of $p$ states, and gray additional pairs involving $s$, $p$, and $d$ states.
(b) At 4 µm internuclear distance and strong dipolar coupling, the varying Förster defects cause avoided crossings of the potentials.
The colorcode in (b) indicates coupling strength to the clock state.
}
\end{figure}

The multi-channel character of the different Rydberg series in \Sr~gives rise to diverse external field responses depending on the principal and orbital quantum numbers.
High level densities and the requirement for magnetic field strengths sufficient to lift the degeneracy in the ground (${}^1S_0$) or clock (${}^3P_0$) manifold, which is often the case for quantum technology applications, can lead to non-generic level shifts and avoided crossings in the Rydberg spectra of single atoms and pairs.

Fig.~\ref{fig:magnetic} shows magnetic field maps for the Förster resonant pair and surrounding states (a) asymptotically in the non-interacting limit and (b) at 4 µm internuclear spacing.
Here, the magnetic field is aligned with the internuclear axis such that the total projection of the hyperfine state is still a conserved quantity and only the $M=-9$ symmetry sector is shown.
For the $5s64s$ equal-state pair, highlighted in black in Fig.~\ref{fig:magnetic}(a), the linear magnetic field response is weak, due to a strong contribution from the singlet channel in that state, and the quadratic response due to diamagnetism dominates even at relatively small field strength.
For such states that are field-seeking, i.e.~they lower their energy with respect to the magnetic field strength due to Zeeman interaction, there exists a local minimum in the magnetic field map, where the pair state is insensitive to magnetic field fluctuations due to a balance between attractive Zeeman and repulsive diamagnetic interaction.
For the state of interest this occurs at approximately $B=50$ G.

Multiple pair states with $p$ character, which are close in energy at zero field, are shifted to smaller energies due to their strong linear response, i.e.~Zeeman effect.
Note that most of the states shown in Fig.~\ref{fig:magnetic} are field seeking because of $M$ being negative.
Only in the case of $f<I$, such states are shifted to larger energies.
However, some pairs of $p$ states, while initially being shifted out of resonance for small fields, become resonant with the $s$ pair again at $B>150$ G.
This is striking and highlights the importance of diamagnetism:
If only linear shifts were considered, all $p$ pairs would be shifted out of resonance, which would drastically change the character of the pair interaction potentials.

At short internuclear distances in Fig.~\ref{fig:magnetic}(b), dipole-dipole interactions are additionally relevant, and the diverse magnetic field responses of the pair states leads to non-trivial functional behavior of interaction potentials, with a multitude of avoided crossings, and of their dipole addressability.
In particular, at $B=200$ G, one of the $p$ pairs initially resonant at zero field, becomes degenerate with a different pair state.
This pair, which is detuned by approximately $-200$ MHz at zero field (cf.~subfigure (a)), is from the same $5sns$ $f=\frac{9}{2}$ Rydberg series, however, for different $n$, namely $n'=n\pm2$, and one of the two states has slightly different singlet-triplet configuration due to its multi-channel nature.
In a magnetic field, this pair remains approximately equidistant to the Förster resonant state.
However, at $B=200$ G, the two $s$ pairs become strongly coupled:
Essentially, two Förster resonances occur simultaneously at approximately 200 MHz energy splitting and the involved $p$ states mediate coupling between the $s$ pairs.

\begin{figure}
\raggedleft
\includegraphics[width=0.51\textwidth]{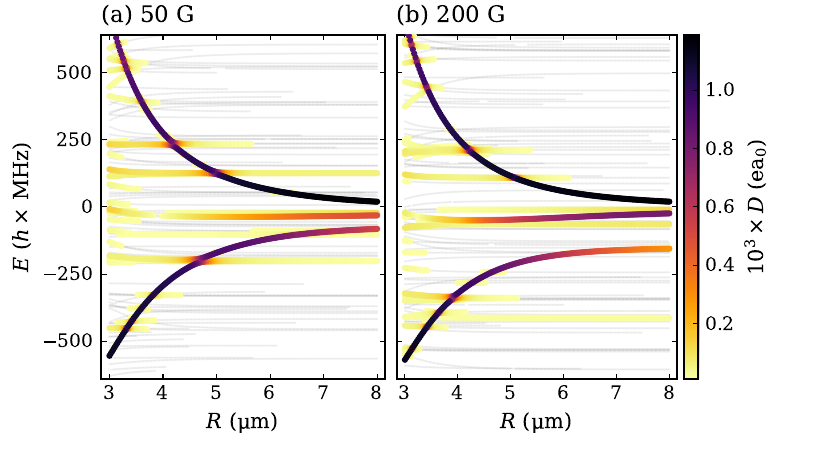}
\caption{\label{fig:competition}
Competition of Förster resonances between pairs $64s+64s$, $63p+64p$, and $62s+66s$.
Their Förster defects depend on the magnetic field strength.
(a) At $B=50$ G, the $64s+64s$ and several $63p+64p$ pairs are close to resonance, while $62s+66s$ is detuned by $\sim 200$ MHz.
(b) At $B=200$ G, both $s$ pairs are resonant with different $p$ pairs while their respective splitting is approximately maintained.
However, the $p$ pairs mitigate a strong coupling between the $s$ pairs, hence the large avoided crossing and dipole addressability.
}
\end{figure}

The corresponding pair potentials are shown in Fig.~\ref{fig:competition} for $B=50$ G (a) and $B=200$ G (b).
Dipolar interactions exceed the energy scale of the asymptotic splitting and the pair interaction potentials feature a broad avoided crossing (cf.~Fig.~\ref{fig:competition}(b)) giving rise to an essentially flat dipole-coupled potential over the full range of $R$.
While for blockade experiments, this situation needs to be avoided, such a potential is highly unusual and may lead to facilitation, where Rydberg pairs are addressed resonantly.

This example illustrates that the magnetic field can effectively act as a knob for competing Förster resonances.
The potentials in Fig.~\ref{fig:competition} can be qualitatively reproduced by a toy model that only includes the four properly symmetrized pair states that contribute dominantly:
\begin{equation}
    \{ \ket{nsns}, \ket{(n-1)pnp}, \ket{n'sn''s}, \ket{(n-1)p'np'} \},
\end{equation}
where only $s$ and $p$ pairs are coupled, respectively, by dipolar interaction (of similar strength) and two different Förster defects control the asymptotic detunings as functions of the magnetic field.

\subsection{Non-perturbative polarizability calculations \label{subsec:electric}}

\begin{figure}
\raggedleft
\includegraphics[width=0.51\textwidth]{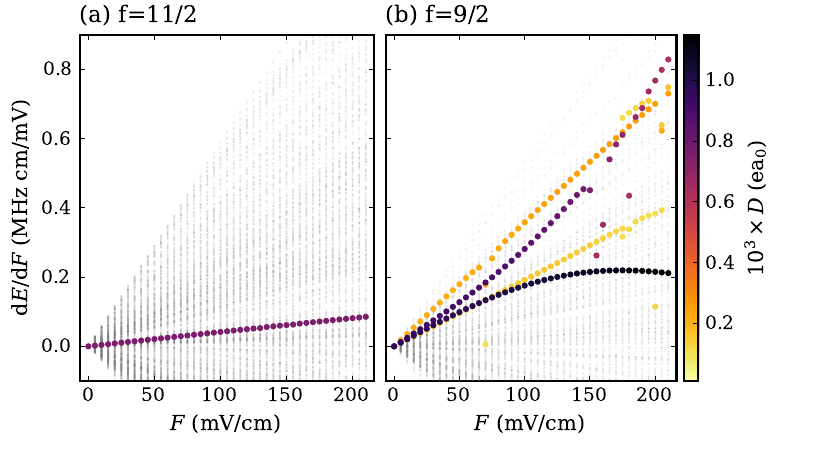}
\caption{\label{fig:polarizabilities}
Electric field response of $5s64s$ equal-state pairs in different Rydberg series at internuclear spacing $R=6$ µm.
(a) The single channel $f=\frac{11}{2}$ state follows the expectation from perturbation theory with linear behavior of the electric field derivative.
(b) For the multi-channel $f=\frac{9}{2}$ state, hyperpolarizability is relevant even at small electric fields, giving rise to configurations with zero polarizability (here: zero local slope).
At resonance, the polarizability of the $f=\frac{9}{2}$ pairs is larger than for $f=\frac{11}{2}$ pairs because of significant $p$ state admixture.
}
\end{figure}

After considering the effects of external magnetic fields in the previous section, we now turn our attention towards electric fields.
Generally, electric fields pose a computational challenge because contributions from high-orbital-angular-momentum states become relevant even for moderate field strengths.
Additionally, the influence of electric fields steeply scales with the principal quantum number ($\sim n^5$).
Therefore, here we restrict our analysis to electric field strengths of few hundreds of mV/cm.
We find that even these small electric fields can cause responses not expected from perturbation theory.

This is illustrated in Fig.~\ref{fig:polarizabilities} for $R=6$ µm internuclear spacing and electric fields aligned with the internuclear axis.
While pairs of the pure triplet state $5s64s$ $f=\frac{11}{2}$ show the expected linear response of the derivative of its eigenenergy with respect to the electric field strength d$E$/d$F$ as a function of the electric field strength $F$, the behavior of the Förster resonant pairs $5s64s$ $f=\frac{9}{2}$ indicates the presence of a hyperpolarizability that becomes relevant for relatively small field strengths $F\sim 150$ mV/cm.
In fact, this can be exploited to achieve extremely high degrees of line stability, when the electric field is tuned to a local extremum of d$E$/d$F$.
Overall, the $f=\frac{9}{2}$ pairs show a much larger polarizability than the $f=\frac{11}{2}$ pairs.
This can be attributed to the strong $p$-state admixture at resonance.
Asymptotically, the pure $5s64s$ $f=\frac{9}{2}$ has near zero polarizability, approximately 2 orders of magnitude smaller than the $5s64s$ $f=\frac{11}{2}$ state.

\subsection{Single channel Rydberg series \label{subsec:11/2}}

\begin{figure}
\raggedleft
\includegraphics[width=0.51\textwidth]{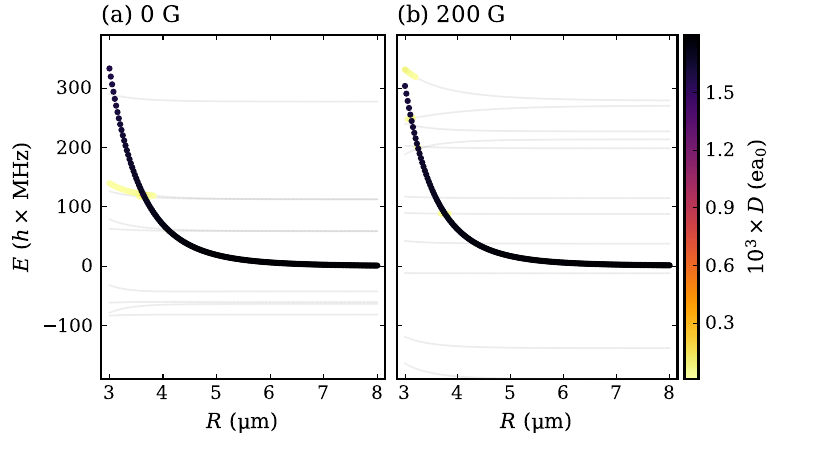}
\caption{\label{fig:11}
Pair potentials for the $5s64s$ $f=\frac{11}{2}$ equal-state pairs for finite magnetic fields $B=0$ G (a) and $B=200$ G (b).
These potentials are spherically symmetric and the characterizing C6 coefficient is only about 5\% smaller in (b) compared to (a).
}
\end{figure}

As mentioned previously, the $5sns$ $f=\frac{11}{2}$ Rydberg series in \Sr~provides, in a lot of aspects, less complexity than the $f=\frac{9}{2}$ series.
It is a pure triplet series, well described in a single channel approximation and, hence, shows a lot of the regular scaling behaviors with respect to the principal quantum number expected from alkali metals or other single-channel Rydberg series.
In this series, an equal-state Förster resonance is not expected and the polarizability scales as $n^7$, while induced-dipolar interaction scales as $n^{11}$.

Fig.~\ref{fig:11} shows the pair interaction potentials (a) at zero magnetic field and (b) at $B=200$ G aligned with the internuclear axis.
Here, the $M=-11$ symmetry sector is shown and the coupling to the clock state is achieved by $\sigma_+$ driving.
Essentially, the magnetic field does not have a strong impact on the potential.
At $B=200$ G the C6 coefficient is reduced by approximately 5\% and the spherically symmetric structure of the potential is maintained (not shown in the figure).
However, to achieve, e.g., Rydberg blockade for pairs within that series, still care must be taken for the choice of the principal quantum number in order to avoid accidental resonances.

\section{Conclusion and Outlook \label{sec:conclusion}}

We have analysed the structure of Rydberg pair-interaction potentials of \Sr, a multi-channel, divalent atomic species.
The ionic core's hyperfine structure gives rise to intriguing features within the Rydberg series, particularly states with vanishing polarizability and resonant dipolar interaction.
While these properties make those states very promising candidates for application in quantum technology, the arising complexity requires careful consideration of external field configurations.
We have shown that magnetic fields can undermine Rydberg blockade because of competing Förster resonances that give rise to avoided crossings in the potential energy landscape and to unanticipated potential shapes, such as an almost completely flat potential with respect to the internuclear distance.
On the other hand, configurations can be found that offer insensitivity to electric fields with pair states that have zero polarizability.

The results from pair-interaction calculations are readily available as input for error models and fidelity estimations in the context of entangling gates in quantum computation \cite{Pagano2022}.
Specifically, the results provide quantitative information about leakage into doubly excited states.
Depending on the available channels, this undesired behavior can potentially be mitigated by appropriate pulse shaping.

We have integrated the tools to analyze the spectra of multi-channel divalent atoms based on quantum defect measurements into the \emph{pairinteraction} software.
Due to the increased level density of fermionic species, the calculation of pair potentials in arbitrary external field configurations remains challenging, particularly for \Sr~due to its large nuclear spin.

Recently, \Yb~has emerged as an interesting candidate for application in quantum computation.
With a nuclear spin of $I=\frac{1}{2}$ it is a natural choice for nuclear qubit encoding.
Due to a similar multi-channel structure, we expect the Rydberg spectrum of \Yb~to share a lot of features with \Sr.
Quantitative predictions have been enabled by related work including spectroscopy of the ${}^1P_1$ and ${}^3P_J$ series in \Yb~ \cite{Peper2024}, which we became aware of during the compilation of this manuscript.

\begin{acknowledgments}

S.W.~and J.M.~acknowledge funding by the Federal Ministry of Education and Research (BMBF) under the Grants QRydDemo, MUNIQC-Atoms, and by the Horizon Europe program HORIZON-CL4-2021-DIGITALEMERGING-01-30 via the project 101070144 (EuRyQa).

\end{acknowledgments}

\bibliography{sr-mqdt}

\end{document}